\documentclass[final,5p,times,twocolumn]{elsarticle}

\usepackage{amssymb}

\usepackage{graphicx,graphics,wrapfig,rotating}         
\usepackage{dcolumn}                            
\usepackage{times,euscript,oldgerm}              %
\usepackage[english]{babel}          
\usepackage{psfrag}

\usepackage{mathtools} 
\usepackage{mathrsfs}
\usepackage{cuted}

\usepackage{natbib}
\usepackage[usenames,dvipsnames]{color}
\usepackage[dvipsnames]{xcolor}
\usepackage[normalem]{ulem}
\usepackage{hyperref}     
\usepackage{stackrel} 
\usepackage{amsmath}
\usepackage{amsthm} 

\usepackage{orcidlink} 
\usepackage{mathbbol} 

\newtheorem{theorem}{Theorem}
\newtheorem{lemma}[theorem]{Lemma}
\newtheorem{definition}[theorem]{Definition}

\newcommand{\ket}[1]{\ensuremath{|#1\rangle}}
\newcommand{\bra}[1]{\ensuremath{\langle #1|}}

\newcommand{\be}{\begin{equation}}
\newcommand{\ee}{\end{equation}}
\newcommand{\ba}{\begin{eqnarray}}
\newcommand{\ea}{\end{eqnarray}}

\newcommand{\Lambdamid}{V_{21}}
\newcommand{\openone}{\mathbb{1}}

\DeclareMathOperator{\Tr}{Tr}

\journal{Physics Letters A}

\begin{document}

\begin{frontmatter}



\title{Distillation of quantum non-Markovianity}


\author[inst1]{Thiago Melo D. Azevedo \orcidlink{0000-0002-1068-1618}}

\affiliation[inst1]{organization={Centro de Informática, Universidade Federal de Pernambuco},
            city={Recife},
           postcode={50.740-560}, 
            state={PE},
            country={Brazil}}

\affiliation[inst2]{organization={Instituto de Física, Universidade Federal da Bahia, Campus de Ondina},
            addressline={Rua Barão do Geremoabo, s.n.}, 
            city={Salvador},
            postcode={40210-340}, 
            state={BA},
            country={Brazil}}

\affiliation[inst4]{organization={Fundação Maurício Grabois},
            addressline={R. Rego Freitas, 192 - República}, 
            city={São Paulo},
            postcode={01220-010}, 
            state={SP},
            country={Brazil}}

\author[inst2,inst4]{Cristhiano Duarte \orcidlink{0000-0001-7267-1477}}
\author[inst3]{Nadja K. Bernardes \orcidlink{0000-0001-6307-411X}}
\affiliation[inst3]{organization={Departamento de Física, Centro de Ciências Exatas e da Natureza, Universidade Federal de Pernambuco},
            city={Recife},
            postcode={50670-901}, 
            state={PE},
            country={Brazil}}

\begin{abstract}
Non-Markovianty of open quantum systems dynamics is a physically relevant property which is usually associated with the backflow of (quantum) information. Using this paradigmatic marker, we develop an operational framework to investigate how non-Markovianity for qubit dynamics can be distilled when many copies of the channels are used, possibly allowing for a stronger effect on the backflow of information.
\end{abstract}



\end{frontmatter}


\section{Introduction}
\label{sec:intro}
Realistic physical systems are never in perfect isolation. The temporal evolution of open quantum systems has been extensively studied and is, given certain constraints, sufficiently well-described via Markovian dynamics generated by a memoryless interaction of a quantum system with its environment~\cite{breuer}. However, strong system-environment interactions, environment correlations, and initial system-environment correlations may cause memory effects rendering the dynamics non-Markovian \cite{Rivas14,Vega17}. Recently, with the advancement of noisy intermediate-scale quantum technologies, there has been considerable interest in describing the dynamics of open quantum systems, since it is crucial for the processing of quantum information and for a better understanding and control of decoherence, decaying, and dissipative effects~\cite{matsuzaki, verstraete,vasile,chin,bogna}.

As thoroughly studied as the concept may be, a unique definition of Markovian dynamics is yet to be established{---and a single definition pinpointing the distinctive features of this class of quantum dynamics is still open for debate. The divisibility of quantum maps appears to be the most used definition \cite{Rivas14, wolf1}, nonetheless, the term is similarly used for maps that satisfy the semi-group property \cite{Alicki06} or for dynamics that do not present a backflow of information \cite{Milz19}. If that were not enough, different degrees of non-Markovianity can be identified, so that these memory effects appear in different forms, which turns the situation even more tortuous. That said, some tasks cannot be achieved with Markovian evolutions, but that can be so if we are allowed some degree of non-Markovianity—improving the performance of quantum heat engines \cite{Thomas18}, enhancing quantum control \cite{Reich15}, and allowing the perfect teleportation of mixed states \cite{elsi}. In this sense, non-Markovianity appears to be a resource for the processing of quantum information.

Following these lines, there has been an attempt to construct a resource theory for quantum processes \cite{berk2021}. Regardless of its generality, recall that one important task of resource theories is to establish whether the resource in question can be distilled. The idea of resource distillation appeared for the first time in the field of quantum information as entanglement distillation \cite{bennett}. Entanglement distillation is a process that extracts high-quality entangled states from a larger number of lower-quality entangled states. The goal is to improve the fidelity and purity of entangled states, which are critical resources for quantum communication and computation protocols. In entanglement distillation, lower-quality entangled pairs are subjected to a series of local operations and measurements, and the resulting data is used to probabilistically select higher-quality pairs. The technique has been widely studied and has important applications in quantum information processing \cite{dur, gisin}.

Elaborating on the question of finding a meaningful definition for Markovian dynamics in a resource theoretical manner, in this paper, we show how non-Markovianity can be distilled when many copies of a channel are available. For this, first, we prove that a Markovian channel cannot create non-Markovianity, in terms established here. Second, the distillation of non-Markovianity will be characterized by two figures of merit: (i) the distinguishability of two states and (ii) the negative eigenvalue of the dynamical matrix. To illustrate our novel operational framework, we investigate a specific qubit non-Markovian channel.

The idea of distilling distinguishability \cite{Wilde19a,Wilde19b} and quantum noisy channels \cite {Regula} was already proposed in different works. However, none of them were interested in non-Markovian effects. Moreover, the advantage of the scheme proposed here is that it is simple enough to have direct experimental applications. Although not interested in distillation, activation of non-Markovianity using high-order maps has been proposed in Ref.~\cite{Maity}. Besides this, there has been a recent effort to define a Markovian resource theory \cite{Anand, Bhattacharya20}. Nevertheless, their scheme is limited to a convex set of Markovian (non-Markovian dynamics), and none of them addresses distillation of non-Markovianity. Non-Markovianity has also been proved to be a quantum dynamical resource in the context of noise reduction and dynamical decoupling \cite{Berk2}. A virtual distillation scheme has been also applied for non-Markovian distillation in Ref.~\cite{Takagi}. However, the proposed protocol is substantially different than the one presented here. Their scheme requires classical postprocessing, and the non-Markovian distillation has been studied in the context of many rounds of implementation.

The paper is organized as follows: section \ref{sec:settings} and \ref{sec:framework} reviews the formalism of quantum evolutions and Markovianity and presents the framework used for this work; section \ref{sec: NM-distillation} describes the process of the distillation of non-Markovianity along with numerical results showing the increase in the distinguishability for the distillation of $2$, $3$ and $4$ copies; section \ref{sec: eigenvalues} contains results for the numerical analysis of the eigenvalues of the intermediate map of the non-Markovian quantum dynamic.

\section{Quantum evolutions}
\label{sec:settings}

We consider throughout this contribution a quantum system $\mathcal{S}$ with associated Hilbert spaces $\mathcal{H}$ isomorphic to some $\mathbb{C}_d$, with an appropriate $d \in \mathbb{N}$. Fixing an orthonormal basis $\{e_1, . . . , e_d\}$
in $\mathcal{H}$ any linear operator in $\mathcal{H}$ may be identified with an $d\times d$ complex matrix, i.e. an element from $M_d(\mathbb{C})$. A mixed state of such a system is represented by a
density matrix, i.e. a matrix $\rho$ from $M_d(\mathbb{C})$ such that $\rho\geq 0$ and $\Tr{\rho}=1$. The identity operator is denoted by the symbol $\openone$. Similarly, the channel $\mathcal{I}:M_d(\mathbb{C})\to M_d(\mathbb{C})$ mapping $\rho$ onto $\rho$ is the identity map. 

We say that a quantum channel, or a dynamical map $\Lambda_t:M_d(\mathbb{C})\to M_d(\mathbb{C})$ is \emph{completely positive} (CP) whenever $(\mathcal{I}\otimes\Lambda_t)(M^{+}_{d^2}(\mathbb{C}))\subset M^{+}_{d^2}(\mathbb{C})$, where $M^{+}_{d^2}(\mathbb{C})=\{A\in M_{d^2}(\mathbb{C})| A \geq 0\}$ is a convex subset of $d^2\times d^2$ positive matrices. We say that a dynamical map $\Lambda_t:M_d(\mathbb{C})\to M_d(\mathbb{C})$ is \emph{trace-preserving} whenever $\mbox{Tr}[\Lambda_{t}(A)]=\mbox{Tr}(A)$, for every $A\in M_{d^2}(\mathbb{C})$.

A general \emph{quantum evolution} is determined by a family $\{\Lambda_{t}\}_{t \geq 0}$ of completely positive trace-preserving maps $\Lambda_t:M_d(\mathbb{C})\to M_d(\mathbb{C})$ such that for $t=0$, $\Lambda_0=\mathcal{I}$. We call the collection of all CPTP time evolutions $\mathsf{CP}$.

We adopt  the usual definition of Markovianity in terms of divisibility into CP processes~\cite{Rivas14, wolf1}: 
\begin{definition}[Markovian dynamics] 
\label{def:Mark_dyn}
We say that a quantum evolution $\Lambda$ is Markovian whenever it is given by CPTP dynamical maps $\Lambda_t$ that are \emph{divisible} in other CP-maps, i.e. 
\begin{equation}
\label{def:divisibility}
\Lambda=\{\Lambda_t\}_{t \geq 0}\,\,\,\text{where}\,\,\,\Lambda_{t}=V_{t,s}\Lambda_{s},
\end{equation}
with $V_{t,s}$ CPTP maps for all times $t\geq s\geq 0$.
\end{definition}
\noindent In turn, any quantum evolution that does not satisfy this condition is called \emph{non-Markovian}. We denote the set of all Markovian time evolutions by $\mathsf{M}\subset\mathsf{CP}$ and non-Markovian time evolutions by $\mathsf{NM}\subset\mathsf{CP}$.

We will also be interested in linear maps $\mathcal{T}: \mathsf{CP}\to \mathsf{CP}$ mapping quantum evolutions into quantum evolutions. We will refer to any such maps as a \emph{dynamical transformation}. Dynamical transformations describe the modifications in a system's dynamics due to physical changes in the system, its environment, and/or the type of interaction between them.

This work aims to investigate whether non-Markovianity can be distilled in the following sense: whether many copies of a non-CP-divisible family of quantum channels $\{ \Lambda_{t} \}_{t \geq 0}$ will result in a stronger effect on the backflow of information, as represented in Fig.~\ref{boxes}. It will become clearer later, but for now, we emphasize that the resulting state $\rho'=\mathcal{T} (\Lambda_t \otimes ... \otimes \Lambda_t)(\rho)$ is also an element from $M_d (\mathbb{C})$. To do so,  we have to show that the dynamical transformation $\mathcal{T}$ preserves Markovian evolutions—put another way, when we are dealing exclusively with Markovian time evolutions, we must guarantee that non-Markovian effects will not appear due to the dynamical transformation's mere presence.

\section{The operational framework}
\label{sec:framework}
Our first step is thus to find a set of Markovianity-preserving dynamical transformations that, nonetheless, will still be meaningful in our operational framework.

\begin{definition}[CP process composition] 
\label{def:CP_pros_comp}
We say that a dynamical transformation $\mathcal{T}: \mathsf{CP}\to \mathsf{CP}$ is a \emph{CP-process composition} whenever, for any $\Lambda\in\mathsf{CP}$, it holds that
\begin{equation}
\label{eq:CP_pros_comp}
\Lambda' \coloneqq\mathcal{T}(\Lambda)=\{\lambda_{t}\, \circ \Lambda_{t}\}_{ t\geq 0},
\end{equation}
where $\lambda_{t}$ is a CPTP map for all $t\geq 0$ .
\end{definition}

In other words, a dynamical transformation $\mathcal{T}: \mathsf{CP}\to \mathsf{CP}$ is a CP-process composition when the net effect of applying $\mathcal{T}$ on an indexed family of quantum maps $\{\Lambda_{t}\}_{t}$ is a possibly index-dependent \textit{displacement} of each element of that family: $\Lambda_{t} \mapsto \lambda_{t} \circ \Lambda_{t}$---where each displacement factor $\lambda_{t}$ is also a CPTP map. Below we prove that the CP process is well-defined.

\begin{lemma}[CPTP preservation] 
\label{theo:CPTP_preserv}
Every CP process composition maps $\mathsf{CP}$ into itself. That is, for all $\Lambda\in\mathsf{CP}$, it holds that $\Lambda'\in\mathsf{CP}$, with $\Lambda'=\{\lambda_t \circ \Lambda_t\}$ for $t \geq 0$.
\end{lemma}

\begin{proof} 
Every CPTP map $\Lambda_t$ has an operator sum representation in terms of its Kraus operators $K_{\alpha}(t)$ \cite{Choi, Kraus}, $\Lambda_t(A)=\sum_{\alpha}K_{\alpha}(t)AK^{\dagger}_{\alpha}(t)$ for $A\in M_d(\mathbb{C})$ and $\sum_{\alpha}K_{\alpha}^{\dagger}(t)K_{\alpha}(t)=\openone_d$. The same is true also for each $\lambda_{t}$, and in this case, assume that their Kraus decompositions are determined by $\{K_{\beta}^{\prime}(t)\}_{\beta}$. Thus, $\lambda_t\circ \Lambda_t(A)=\sum_{\alpha,\beta}K'_{\beta}(t)K_{\alpha}(t)AK_{\alpha}^{\dagger}(t)K_{\beta}^{\prime \dagger}(t)=\sum_{\gamma}K^{\prime \prime}_{\gamma}(t)XK^{\prime \prime \dagger}_{\gamma}(t)$, where we define $K_{\gamma}^{\prime \prime}=K_{\beta}^{\prime}K_{\alpha}$ with $\gamma=(\alpha,\beta)$, and in this case $\sum_{\gamma}K^{\prime \prime\dagger}_{\gamma}(t)K^{\prime \prime}_{\gamma}(t)=\openone_d$. 
\end{proof}

Although CP-process composition is guaranteed to take CP maps onto CP maps, the same cannot be said about Markovianity---unless we demand additional conditions. The structure of divisibility imposed by CPTP-divisibility does not easily carry over to the composition $\lambda_{t} \circ \Lambda_{t}$ even if $\{\Lambda_{t}\}_{t}$ is divisible. To see that this is the case, suppose that $\{V_{t,s}\}_{(t,s)}$ are the `intermediate maps' for the CPTP-divisible family $\{\Lambda_{t}\}_{t}$, and that $\Lambda^{\prime}_{t}=\lambda_{t} \circ \Lambda_{t}$ are the displaced maps we are investigating the potential CPTP-divisibility. Indeed, for every $t \geq s \geq 0:$
\begin{align}
    \Lambda_{t}^{\prime} = \lambda_{t} \circ \Lambda_{t} = \lambda_{t} \circ V_{t,s} \circ \Lambda_{s} = V_{t,s}^{\prime} \circ \Lambda_{s}.
\label{Eq.WhereCPTPDivFails}
\end{align}
Comparing Eqs.~\eqref{def:divisibility} and \eqref{Eq.WhereCPTPDivFails}, we note that although there is a well-defined CPTP `intermediate map' connecting two consecutive instants of time in the latter---so that the functional relation of that equation is correct---the dynamical maps on either side of the equality do not have the same nature. On the left-hand side of Eq.~\eqref{Eq.WhereCPTPDivFails} we are dealing with the displaced map, whereas on the other side, we are dealing with the original family. One could read that equation slightly differently, for every $t \geq s \geq 0:$
\begin{align}
\lambda_{t} \circ \Lambda_{t} = V_{t,s}^{\prime} \circ \Lambda_{s}.
\label{Eq.MotivatingModCPTPDiv}
\end{align}

Without assuming anything about the nature of $\{\Lambda_{t}\}_{t}$, other than saying that they obey Eq.~\eqref{Eq.MotivatingModCPTPDiv}, we run into the same issue as before: Eq.~\eqref{Eq.MotivatingModCPTPDiv} is essentially identical to Eq.~\eqref{eq:CP_pros_comp}, but without demanding extra assumptions, they may express different concepts. Either set of equations says something about the divisibility of the family $\{\Lambda_{t}\}$, but expresses it in a different functional form. The next definition and results summarize this discussion.

\begin{definition}[Modified CP-Divisibility] 
\label{def:mod-divisibility}
Let $\{ \Lambda_{t} \}_{t \geq 0}$ be a family of CPTP maps and $\lambda$ be a fixed CPTP map. We say that $\{ \Lambda_{t} \}_{t \geq 0}$ is \emph{mod-CP-divisible} for $\lambda$ whenever:

\begin{equation}
    \forall\, t \geq s \geq 0, \exists \,  V_{t,s}: \lambda \circ \Lambda_{t} = V'_{t,s} \circ \Lambda_{s},
\end{equation}
where $V'_{t,s}$ is CPTP.
\end{definition}

For this modified notion of CP-Divisibility, where essentially there is an extra fixed decoder map $\lambda$ on the left-hand side,}  we can associate yet another definition of Markovianity. Our argument is inspired by the discussion found in Ref.~\cite{Rivas14}.

\begin{definition}[Modified Markovianity]
\label{def:mod-markovian}
A quantum system subject to a time evolution given by
some family of CPTP maps $\{ \Lambda_{t} \}_{t}$ is Mod-Markovian whenever if, for every $t \geq s \geq 0$, there exists a CPTP map $V'_{t,s}$, such that $\lambda \circ \Lambda_{t} = V'_{t,s} \circ \Lambda_{s}$. That is, if the family  $\{ \Lambda_{t} \}_{t \geq 0}$ is mod-CP-divisible.
\end{definition}

In Ref.~\cite{rivas}, Rivas et. al. established a paradigmatic connection between the usual notion of Markovianity and the temporal non-increasing of the associated 1-norm, represented by $\Vert \cdot \Vert_{1}$. As our modified definition of Markovianity is essentially structurally the same as the standard one, it is expected that we can recover an analogous characterization in terms of the 1-norm. The first result below is a reminder of Rivas et. al.'s result, whereas the subsequent theorem is our adapted contribution. 

\begin{theorem}[Theorem 3.4 in \cite{Rivas14}]
A quantum evolution $\{V_{t,s}, t \geq s \geq 0\}$
is Markovian if and only if for all $t$ and $s$, $t\geq s$:
\begin{align}
\Vert \left( V_{t,s} \otimes \openone \right)(A) \Vert_{1} \leq \Vert A \Vert_{1}
\label{Eq.Characterizing_Usual_Markovianity}
\end{align}
for every Hermitian operator $A$ acting on $\mathcal{H} \otimes \mathcal{H}$, and where the 1-norm $\Vert . \Vert_{1}$ is defined as follows:
\begin{align}
\Vert A \Vert_{1} := \mathrm{Tr} \left(\sqrt{A^{\dagger}A}\right).   
\end{align}
\label{theo:Usual_Markov_Preserving}
\end{theorem}

\begin{theorem}[Mod-Markovian 1-norm non-increasing] \label{theo:Mark_preserv}
A quantum evolution $\{\Lambda_{t}\}_{t \geq 0}$ is Mod-Markovian for $\lambda$, then for every $t \geq s \geq 0$, each intermediate map $V'_{t,s}$ has the property 
\begin{align}
\left\Vert (V'_{t,s} \otimes \openone )({A})  \right\Vert_1 \leq \left\Vert A \right\Vert_1,
\end{align}
for every Hermitian operator $A$ acting on $\mathcal{H} \otimes \mathcal{H}$.
\end{theorem}

\begin{proof}
Since for a mod-Markovian time evolution $\{ \Lambda_{t} \}_{t\geq 0}$, the intermediate map $ V'_{t,s} = \lambda \circ V_{t,s}$ is
completely positive for any $t\geq s \geq 0$, the map $ V'_{t,s} \otimes \openone $ is
positive. Since for every trace-preserving positive map $\Lambda$ and Hermitian operator $A$,  $\left\Vert\Lambda(A)\right\Vert_1 \leq \left\Vert A \right\Vert_1  $, we have our result. 
\end{proof}

\section{Non-Markovian distillation}
\label{sec: NM-distillation}

\begin{figure}[h]
    \centering
    \includegraphics[scale=0.5]{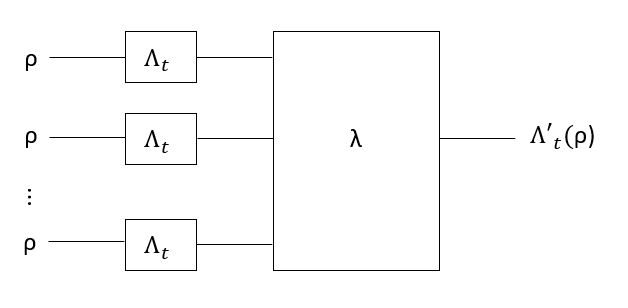}
    \caption{Circuit representation of the proposed quantum process for non-Markovian distillation.} 
    \label{boxes}
\end{figure}

Now that we introduced the operational framework we will be dealing with, we start investigating the possibility of NM distillation. For simplicity, we focus on the possibility of NM distillation for one-qubit channels. Imagine we have $n$ copies of the same non-Markovian channel $\Lambda_t$ and one displacement map $\lambda$, such that $\Lambda_t'=\lambda \circ (\Lambda_t\otimes...\otimes\Lambda_t):M_{d^n}(\mathbb{C})\to M_d(\mathbb{C})$, as represented in Fig.~\ref{boxes}. The main objective of this work is to determine whether the resulting channel $\Lambda_t'$ can be more non-Markovian than the original one.

Establishing the (non-)Markovianity of quantum dynamics efficiently is not an easy task in general. There are many different approaches in the literature to measure, detect, or witness non-Markovianity \cite{Rivas14}. Typically, these quantities are based on the monotonic decay under
the action of CPTP maps. Retrieval of these quantities in time is then connected to a backflow of information from the environment to the system and signals a non-Markovian dynamics.

In this work, we want to connect the non-Markovian feature of the dynamics to backflow of information. Commonly, detecting backflow of information is done by observing a recovery in time of entanglement between a system and an ancilla, as presented in the seminal work of Rivas, Huelga, and Plenio \cite{rivas}. Another possibility is to witness the backflow by observing an increase in the distinguishability $D(\rho_1,\rho_2)$ of two states $\rho_1$ and $\rho_2$, defined by \cite{blp} 
\begin{equation}
D(\rho_1, \rho_2):=\frac{1}{2}\left\Vert \rho_2-\rho_1 \right\Vert_1.
\end{equation}
With this measure, whenever the distinguishability increases in time $D(\Lambda_t(\rho_1),\Lambda_t(\rho_2))>D(\Lambda_s(\rho_1),\Lambda_s(\rho_2))$, for a given pair of states $\rho_1$ and $\rho_{2}$, and $t > s$, this increasing signals that the quantum dynamics is \emph{not} Markovian—we usually say that the inequality is a witness of non-Markovianity. Since in this work the aim is to prove the possibility of distilling non-Markovianity and entanglement is already a resource in quantum information processing by itself, we will restrict our analysis in terms of distinguishability.

Note that distinguishability is a key element in quantum key distribution protocols, where the security is based on the non-orthogonality of quantum states. Moreover, distinguishability is also crucial for bit commitment protocols which are building blocks of a series of protocols such as secure coin flipping, zero-knowledge proofs, secure computation, user authentication, signature schemes, and verifiable secret sharing \cite{salvail}. Distinguishability is also presented as a source for quantum speedup \cite{Vedral10}.

In this work, we measure the degree of non-Markovianity by calculating a temporal change in distinguishability. For a single copy of the channel, it is defined by 

\begin{equation}
\begin{aligned}
         \Delta D(\rho_1,\rho_2) := \frac{1}{2} \left(\left\Vert \Lambda_{t}(\rho_2-\rho_1)\right\Vert_1- \left\Vert \Lambda_{s}(\rho_2-\rho_1)\right\Vert_1\right).
\end{aligned}
\label{Eq.DeltaD}
\end{equation}
Note that in eq.\eqref{Eq.DeltaD} we are essentially comparing the distinguishability between $\rho_1$ and $\rho_2$ against two subsequent time instants, $t > s$. Whenever $\Delta D(\rho_1, \rho_2)$ is positive, this signals that the dynamics is not Markovian. We will assume that the greater that quantity is, the more non-Markovian a particular dynamic is.

We would like to verify if the resulting dynamics $\Lambda'_t$ is more non-Markovian. Given our operational framework, we need to adapt accordingly eq.\eqref{Eq.DeltaD}. The temporal increase in the distinguishability between two states $\rho_1$ and $\rho_2$ for $n$ copies will be defined as 

\begin{equation}
\begin{aligned}
         &\Delta D_n(\rho_1,\rho_2) := \\
         &  \frac{1}{2} \{\left\Vert \lambda \circ (\Lambda_t\otimes...\otimes\Lambda_t)(\rho_2\otimes...\otimes\rho_2-\rho_1\otimes...\otimes\rho_1)\right\Vert_1 \\-&
         \left\Vert \lambda \circ (\Lambda_s\otimes...\otimes\Lambda_s)(\rho_2\otimes...\otimes\rho_2-\rho_1\otimes...\otimes\rho_1)\right\Vert_1\}.
\end{aligned}
\end{equation}
Note that if $\lambda=\mathcal{I}$, then $\Delta D_1(\rho_1,\rho_2)=\Delta D(\rho_1,\rho_2)$.

In our framework, the displacement map $\lambda$ acts as a coarse-graining map. We do so because we will compare final states with the same dimension, as in the protocol of entanglement distillation. Similar to what was defined in Ref.~\cite{Chris-coarse}, any coarse-graining map $\lambda$ that takes $\rho \otimes... \otimes \rho \in M_{d^n}(\mathbb{C})$ as input and has $\lambda(\rho)=\rho' \in M_{d}(\mathbb{C})$ as output can be defined—via the Stinespring dilation—as 
\begin{equation}
    \lambda(\rho)= \text{Tr}_{D,r}[U(\rho\otimes...\otimes \rho \otimes \ket{0}\bra{0}_{r}\otimes\ket{0}\bra{0}_{d})U^{\dagger} ],
\end{equation}
where the last two states $\ket{0}\bra{0}_{r}$ and $\ket{0}\bra{0}_{d}$ belong to auxiliary
spaces $M_{r}(\mathbb{C})$ and $M_{d}(\mathbb{C})$, respectively. Here $d=2$, and $r=2$. The unitary map $\mathcal{U}(\cdot) = U(\cdot)U^{\dagger}$ acts on the space  $M_{d^n}(\mathbb{C})\otimes M_r(\mathbb{C})\otimes M_d(\mathbb{C})$.  Note that this type of transformation satisfies Theorem~\ref{theo:Mark_preserv}. A detailed proof of it can be found in \ref{sec:ModMarkovianityInProposedProtocol}.

For $\Lambda_t$, we used the dynamical map presented in Ref.~\cite{nadja-NM}, where a collisional model describes the evolution. Quantum collisional models are depicted by repeated interactions between an open quantum system and particles of the environment ($\omega_{env}$). Each interaction between the system and the $j$-th particle of the environment is represented by unitary $U_j$ lasting for a fixed time $\tau$. The particles of the environment are typically identical and noninteracting. The resulting evolution after $j$ collisions is given by $\rho(j\tau)=\text{Tr}_{j}...\text{Tr}_{1}[U_j...U_1(\rho(0)\otimes \omega_{env})U_1^{\dagger}...U_j^{\dagger}]$. This simulates a Markovian dynamics. For non-Markovian dynamics, it has been shown in Ref.~\cite{nadja-NM} that the particles of the environment can no longer be noninteracting, but they have to present some degree of correlation. This will be encoded here by a parameter $\epsilon$. We are then interested in three different instants of time, the initial time $0$, the intermediate time $s$, and the final time $t$. Thus, just two collisions are enough for the purpose of this work with $s=\tau$ and $t=2\tau$. Since the $\tau$ is fixed, we will omit it from the notation. The channels for both the middle time and the last time are given as

\begin{equation}
    \label{NM Time Evolution}
    \begin{split}
    \Lambda_{1}(\rho) =& (1-2\epsilon)\rho +\epsilon(\sigma_Z \rho \sigma_Z + \sigma_X\rho  \sigma_X),\\
    \Lambda_{2}(\rho) =& (1-2\epsilon)^2\rho+4\epsilon^2\rho+2\epsilon(1-2\epsilon)(\sigma_Z \rho \sigma_Z 
    + \sigma_X\rho\sigma_X).
    \end{split}
\end{equation}
 The advantage of this model is that the parameter $\epsilon$ ($0\leq \epsilon \leq 0.5$) is related to the strength of the non-Markovianity. When $\epsilon>0.25$, we are in a strong non-Markovian regime, while for $\epsilon \leq 0.25$ we are in a weak non-Markovian regime \cite{nadja-NM, sabrina}. These regimes are defined by the intermediate map $V_{t,s}$. If the intermediate map is given by a positive (but not CP) map, the dynamics are called weak NM. Otherwise, if the intermediate map is not even positive, the evolution is referred to as strong NM. It is possible to show that strong NM always presents a backflow of information in terms of distinguishability \cite{Dariusz15}.

Notice that the analyzed model belongs to the collisional model class (for a review, see \cite{CICCARELLO20221}). Although we are dealing only with a specific example, this example could be easily generalized for any unital map. Moreover, for qubits, collisional models can simulate any non-Markovian dynamics \cite{Rybar}. Our aim here is to find an example where we could prove that the proposed distillation scheme works. However, the scheme can be easily applied to other examples.

Time is discretized in this example, as in all collisional models. For the work here this is not a problem because the idea was to show an increase in non-Markovianity between two time steps. However, the protocol proposed is quite general and, in principle, could work for other continuous in-time dynamics.

\subsection{Two copies}
\label{sec:2_copies}

Let us start with a distillation scheme using two copies of the proposed NM channel $\Lambda_t$. First, we want to find a displacement map $\lambda$ that gives an increase in the distinguishability, consequently, showing an improvement in the non-Markovianity. Thus, the quantity that we will analyze is $\Delta D_2(\rho_1,\rho_2)$.

For simplicity, we impose a restriction that our initial states $\rho_1$ and $\rho_2$ are pure and orthogonal states, where $\rho_1$ is given by:
\begin{gather}\label{eq:pure_orthogonal_1}
\rho_1=\frac{1}{2}
\begin{bmatrix}
1+\cos{\theta} & \sin{\theta} \cos{\phi}+i\sin{\theta}\sin{\phi}\\
\sin{\theta} \cos{\phi}-i\sin{\theta}\sin{\phi} & 1-\cos{\theta}
\end{bmatrix},
\end{gather}

and $\rho_2$ is given by:
\begin{gather}\label{eq:pure_orthogonal_2}
\rho_2=\frac{1}{2}
\begin{bmatrix}
1-\cos{\theta} & -\sin{\theta} \cos{\phi}-i\sin{\theta}\sin{\phi}\\
-\sin{\theta} \cos{\phi}+i\sin{\theta}\sin{\phi} & 1+\cos{\theta}
\end{bmatrix},
\end{gather}

with $0\leq\theta\leq\pi$ and $0\leq\phi\leq 2\pi$. Since the objective is to find at least one case of NM distillation, this restriction is not an issue. Note that it has been proved that this choice of state optimizes the recovery of distinguishability \cite{Wissmann}.

As we explained, the displacement map $\lambda$ will be considered a coarse-graining map. Thus, to optimize this operation, we need to choose a proper unitary operator, $U$, with the appropriate input and output dimensions. To simplify the calculations, we opted to restrict $U$ to matrices in which all the elements outside the two diagonals are zero.  This was inspired by the class of mixed states known as X-states proposed by Yu and Eberly \cite{Eberly07}. As shown in this reference, this class of states covers a wide range of entangled states. When we choose the unitary operator in the X-form, we guarantee, thus, that we will not neglect at least this type of correlated states, and as we know, how the system and environment correlate is essential for non-Markovianity.

For our matrix $U$, we name the elements in the main diagonal $u_i$ and the elements in the secondary diagonal $v_i$, $i=1,2,...,16$. We calculated $\Delta D$ and $\Delta D_2$ in Eq.~\ref{distinguish before lambda} and~\ref{distinguish after lambda}.

Since our objective is to find one case of distillation of non-Markovianity and not necessarily optimize the distillation, we can further limit our selections of matrices and restrict ourselves to cases where $u_i$ and $v_i$ are either $0$ or $1$. Additionally, maximizing the distinguishability of states is not a simple problem for optimization techniques such as semi-definite programming which leads to difficulties in finding an optimal solution. Since $U^{\dagger}U=1$, we have, for the main diagonal, $u_i^2+v_{17-i}^2=1$, and for the secondary diagonal, $u_iv_i+u_{17-i}v_{17-i}=0$. So, for each line $i$, we must have  $u_i=u_{17-i}\neq v_i=v_{17-i}$. Since we want $\Delta D_2\neq 0$, we have two choices: $u_1 = 1$, which implies $v_1 = 0$,  $u_{16} = 1$, $v_{16} = 0$, $u_{13} = 0$, $v_{13} = 1$, $v_4 = 1$, $u_4 = 0$; or, $u_1 = 0$, which implies $v_1 = 1$,  $u_{16} = 0$, $v_{16} = 1$, $u_{13} = 1$, $v_{13} = 0$, $v_4 = 0$, $u_4 = 1$. The other components do not have any restrictions, so we are free to choose. We ended up with the matrix $U$ given by Eq.~\ref{matrix V}.

\begin{strip}
\begin{equation}
    \label{distinguish before lambda}
    \begin{aligned}
        \Delta D=& \frac{-\sqrt{2}}{2}\Bigl| 2+2 \epsilon (-5 + 7 \epsilon) - 2 \epsilon (-1 + 3 \epsilon) (\cos{2 \theta} +2\cos{2 \phi} \sin^2{\theta}) \Bigr|^{1/2}\\
                                   & +  \Bigl| 1 + 2 \epsilon (-1 + 2 \epsilon) (5 + 14 \epsilon (-1 + 2 \epsilon))\\
                                   & - 2 \epsilon (-1 + 2 \epsilon) (1 + 6 \epsilon (-1 + 2 \epsilon)) (\cos{2 \theta} + 2 \cos{(2 \phi)} \sin^2{\theta}) \Bigr|^{1/2},
    \end{aligned}
\end{equation}

\begin{equation}
    \label{distinguish after lambda}
    \Delta D_2=-\frac{1}{2}\left\{\left|1 - 2 \epsilon\right| - \left|1 - 4 \epsilon + 8 \epsilon^2\right| \right\} \left\{\left|
    |u1|^2 - |u13|^2\right| + 
   \left| |v16|^2  - |v4|^2 \right|\right\} \left|\cos{\theta}\right|.
\end{equation}
\begin{equation}\label{matrix V}
      U=\left(
\begin{array}{cccccccccccccccc}
 1 & 0 & 0 & 0 & 0 & 0 & 0 & 0 & 0 & 0 & 0 & 0 & 0 & 0 & 0 & 0 \\
 0 & 0 & 0 & 0 & 0 & 0 & 0 & 0 & 0 & 0 & 0 & 0 & 0 & 0 & 1 & 0 \\
 0 & 0 & 0 & 0 & 0 & 0 & 0 & 0 & 0 & 0 & 0 & 0 & 0 & 1 & 0 & 0 \\
 0 & 0 & 0 & 0 & 0 & 0 & 0 & 0 & 0 & 0 & 0 & 0 & 1 & 0 & 0 & 0 \\
 0 & 0 & 0 & 0 & 0 & 0 & 0 & 0 & 0 & 0 & 0 & 1 & 0 & 0 & 0 & 0 \\
 0 & 0 & 0 & 0 & 0 & 0 & 0 & 0 & 0 & 0 & 1 & 0 & 0 & 0 & 0 & 0 \\
 0 & 0 & 0 & 0 & 0 & 0 & 0 & 0 & 0 & 1 & 0 & 0 & 0 & 0 & 0 & 0 \\
 0 & 0 & 0 & 0 & 0 & 0 & 0 & 0 & 1 & 0 & 0 & 0 & 0 & 0 & 0 & 0 \\
 0 & 0 & 0 & 0 & 0 & 0 & 0 & 1 & 0 & 0 & 0 & 0 & 0 & 0 & 0 & 0 \\
 0 & 0 & 0 & 0 & 0 & 0 & 1 & 0 & 0 & 0 & 0 & 0 & 0 & 0 & 0 & 0 \\
 0 & 0 & 0 & 0 & 0 & 1 & 0 & 0 & 0 & 0 & 0 & 0 & 0 & 0 & 0 & 0 \\
 0 & 0 & 0 & 0 & 1 & 0 & 0 & 0 & 0 & 0 & 0 & 0 & 0 & 0 & 0 & 0 \\
 0 & 0 & 0 & 1 & 0 & 0 & 0 & 0 & 0 & 0 & 0 & 0 & 0 & 0 & 0 & 0 \\
 0 & 0 & 1 & 0 & 0 & 0 & 0 & 0 & 0 & 0 & 0 & 0 & 0 & 0 & 0 & 0 \\
 0 & 1 & 0 & 0 & 0 & 0 & 0 & 0 & 0 & 0 & 0 & 0 & 0 & 0 & 0 & 0 \\
 0 & 0 & 0 & 0 & 0 & 0 & 0 & 0 & 0 & 0 & 0 & 0 & 0 & 0 & 0 & 1 \\
\end{array}
\right).
\end{equation}
\end{strip}

\begin{figure}[h]
    \centering
    \includegraphics[scale=0.42]{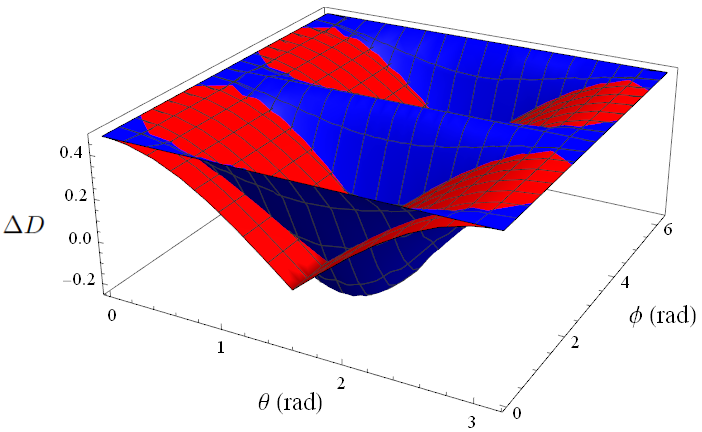}
    \caption{Graph of the variation of distinguishability for 2 copies, with $\epsilon=0.4$. $\Delta D_2$ is in red and surpasses $\Delta D$, in blue, for a range of values of $\theta$ and $\phi$.}
    \label{D 2 copias- 04}
\end{figure}

We plotted $\Delta D$ and $\Delta D_2$ for $\epsilon=0.4$, in Fig.~\ref{D 2 copias- 04}. Note that $\epsilon\geq0.25$ means a regime of strong non-Markovianity. The graphs are functions of the initial states $\theta$ and $\phi$. In Fig.~\ref{D 2 copias- 04}, we see that there are choices for $(\theta,\phi)$ where $\Delta D_2>\Delta D>0$. This is exactly the region we wanted to spot, where the distillation process was successful. However, depending on the choice of $(\theta,\phi)$, $\Delta D_2<\Delta D$. Note here that no optimization was implemented. So it can be that other choices of displacement maps can give a better result.

\begin{figure}[h]
    \centering
    \includegraphics[scale=0.45]{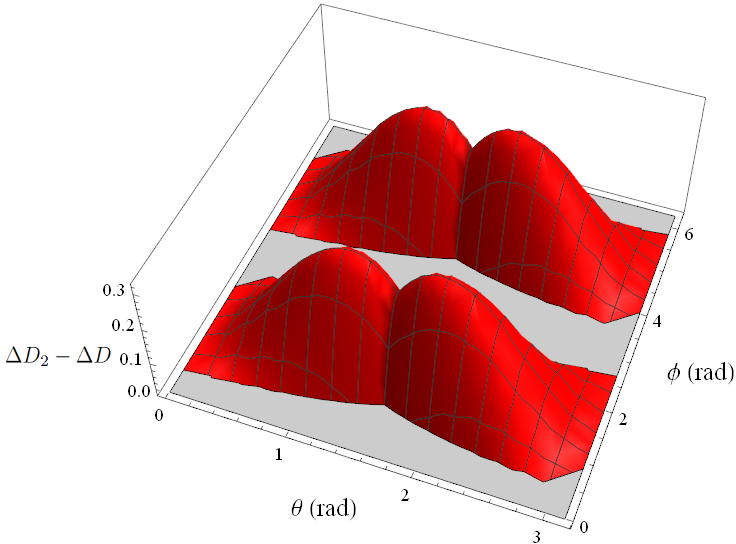}
    \caption{Graph of the difference $\Delta D_2 - \Delta D$ for 2 copies, with $\epsilon=0.4$. Only positive values of the difference are shown, i.e. when $\Delta D_2> \Delta D$, corresponding to which values of $\phi$ and $\theta$ the distillation is successful. The graph is symmetric over the $\theta = \pi/2$ line.} 
   \label{Diff_delta_D}
\end{figure}

In Fig.~\ref{Diff_delta_D}, we plotted the difference $\Delta D_2-\Delta D$ and showed only the positive values of the graph so that we can better identify the regions in which the distillation is successful. We also calculated a global maximum of $\Delta D_2-\Delta D=0.322$, in four points $(\theta, \phi) = [(0.386 \pi, 0.5 \pi), (0.386 \pi, 1.5 \pi), (0.614 \pi, 0.5 \pi), (0.614 \pi, 1.5 \pi)]$. We can see that the graph is symmetric over the $\theta = \pi/2$ line, and the maximum values of $\Delta D_2-\Delta D$ are in the region near $\phi = \pi/2$ and $\phi = 3\pi/2$, with two maximums for each of these values for $\phi$.

A question that needs to be addressed is whether the increase of distinguishability could come simply from using a coarse-graning displacement map $\lambda$ without any dynamics. We look into this situation in \ref{sec:coarse-graining-map}. There, we show that for the example we are working with here, there cannot be an increase of distinguishability arising only from applying the displacement map. Nonetheless, the same may not hold for other cases. This is an inherent feature of our operational framework and should be handled on a case-by-case basis.

\begin{figure}[h]
    \centering
    \includegraphics[scale=0.42]{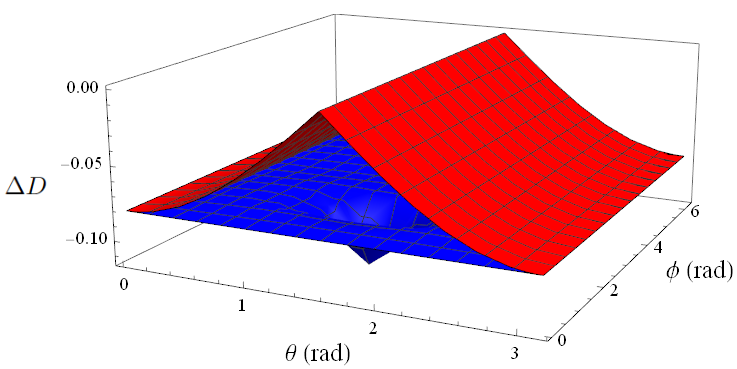}
    \caption{Graph of the variation of Distinguishability for 2 copies, with $\epsilon=0.2$. $\Delta D_2$ is in red and is greater or equal to $\Delta D$ in Blue for all $\theta$ and $\phi$. However, they are both below zero, so this case does not interest us.}
   \label{D 2 copias- 02}
\end{figure}

Finally, we also plotted $\Delta D$ and $\Delta D_2$ for the case of weak non-Markovianity, with $\epsilon=0.2$, in Fig.~\ref{D 2 copias- 02}. Here all the values were negative, so we could not witness the success of the distillation process in this case. However, the dynamics can be non-Markovian even if there is no back-flow of information in terms of the distinguishability of two states.

\subsection{Three and four copies}
\label{sec:3_copies}

\begin{figure}[]
    \centering
    \includegraphics[scale=0.37]{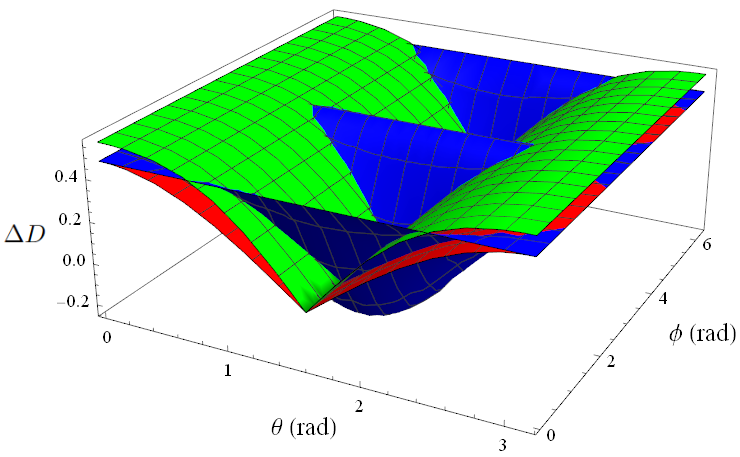}
    \caption{Comparative of $\Delta D$, $\Delta D_2$ and $\Delta D_3$ for $\epsilon=0.4$: in Blue is the original distinguishability, in Red is the case after the distillation of 2 copies and in green after the distillation of 3 copies.}
    \label{distinguibilidade 3 copias}
\end{figure}

We also analyzed the distinguishability for three copies. First, we need to choose a matrix $U$ to be used. For this case, we opted for a matrix $U$ that only has elements in the two diagonals, and these elements are bounded to $[0,1]$. Then, we performed a variation numerically of the entries of $U$ to try to maximize the increase in distinguishability. We obtained the plot of $\Delta D_3$, $\Delta D_2$ and $\Delta D$ in terms of the initial state's parameters $(\theta,\phi)$ in Fig.~\ref{distinguibilidade 3 copias}, with $\epsilon=0.4$. We see that there are regions for $(\theta,\phi)$ where $\Delta D_3>\Delta D_2 > \Delta D>0$. So, the use of more copies was advantageous for the distillation scheme. Note also that there are regions where $\Delta D_3$ is bigger than the maximum value of $\Delta D$ showing a clear advantage in the distillation scheme.
 
For four copies, we obtained the same result as for three copies. Recall that we are not optimizing our results, as we are choosing a very particular form for the unitary $U$, so that one could, in theory, improve the results by finding the unitary that maximizes the difference between the distinguisabilities and the most negative eigenvalues. Nonetheless, even in this crude case, we can see that our operational framework extracts non-Markovianity out of the dynamics.

\section{Analysis of the Eigenvalues of the dynamical matrix}
\label{sec: eigenvalues}

As we observed before, the distinguishability of states is a witness of non-Markivianity, but it will fail to detect all non-Markovian dynamics—put another way, it is not a necessary and sufficient criterion in general. In this way, it will be important to consider an additional way to verify if a map is non-Markovian. For this, we resort to the structural form of CPTP-divisible maps, and we will have to check whether the possible intermediate map $V_{t,s}$ is CP or not. One way to do this is to explore the Choi matrix representation of the map. All the information about a map $\Lambda$ can be extracted from the Choi matrix representation defined by $\mathcal{I}\otimes \Lambda(\ket{\phi_+}\bra{\phi_+})$, where $\ket{\phi_+}$ is a maximally entangled state. This is also known as the Choi-Jamiolkowski isomorphism \cite{Choi, Jamiokowski72}. A map $\Lambda$ is CP iff $\mathcal{I}\otimes \Lambda(\ket{\phi_+}\bra{\phi_+})\geq 0$ \cite{Choi, Bengtsson06}. In this way, it is possible to check if the evolution $\Lambda_t$ is non-Markovian by calculating the most negative eigenvalue of the intermediate map in its Choi form $\mathcal{I}\otimes V_{t,s}(\ket{\phi_+}\bra{\phi_+})$, which will be denoted $\zeta$. 

For a general analysis of the non-Markovian character of the evolution, now we analyze the most negative eigenvalue of the dynamical matrix of the intermediate maps for the different scenarios analyzed before.

\begin{figure}[h]
    \centering
    \includegraphics[scale=0.7]{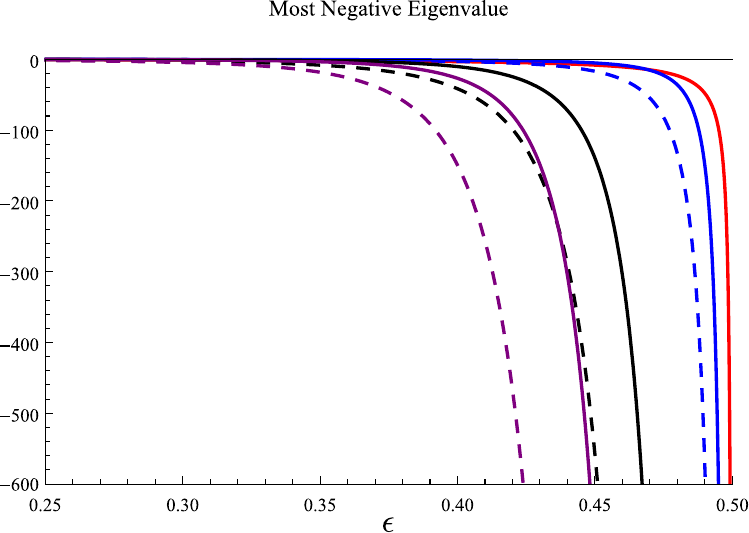}
    \caption{Comparison of the most negative eigenvalues of the dynamical matrices: of $\Lambdamid$ is in red, of   $\Lambdamid\otimes\Lambdamid$ in dashed blue, of $\lambda(\Lambdamid \otimes \Lambdamid)$ in blue, of $\Lambdamid\otimes\Lambdamid\otimes\Lambdamid$ in dashed black, of $\lambda(\Lambdamid\otimes\Lambdamid\otimes\Lambdamid)$ in black,  $\Lambdamid\otimes\Lambdamid\otimes\Lambdamid\otimes \Lambdamid$ in dashed purple, $\lambda(\Lambdamid\otimes\Lambdamid\otimes\Lambdamid\otimes \Lambdamid)$ in purple.}
    \label{autovalores 4 copias}
\end{figure}

From Fig.~\ref{autovalores 4 copias}, it is clear that in all cases we studied, the eigenvalues are negative, which is a witness of non-Markovianity \cite{rivas}. One can also notice that, for a fixed $\epsilon$, the distilled channels have a larger absolute value for the negative eigenvalues than the non-distilled counterparts, while not being greater than the negative eigenvalues for multiple copies. However, in this context, the measure of non-Markovianity is proportional to the area of the eigenvalue which is below zero \cite{rivas}. We have estimated the area determined by the curves in Fig. \ref{autovalores 4 copias}, for all the cases analyzed, and those integrals diverge. Thus, according to this particular quantifier, we cannot conclude that one scenario is more non-Markovian than the other.

\begin{figure}[h]
    \centering
    \includegraphics[scale=0.7]{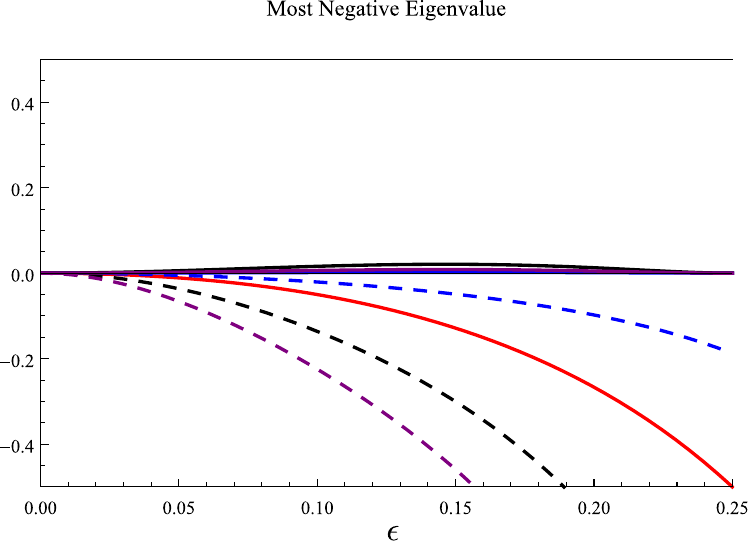}
    \caption{Comparison the most negative eigenvalues for $0.00 \leq\epsilon \leq 0.25$ of the dynamical matrices: of $\Lambdamid$ is in red, of   $\Lambdamid\otimes\Lambdamid$ in dashed blue, of $\lambda(\Lambdamid \otimes \Lambdamid)$ in blue, of $\Lambdamid\otimes\Lambdamid\otimes\Lambdamid$ in dashed black, of $\lambda(\Lambdamid\otimes\Lambdamid\otimes\Lambdamid)$ in black,  $\Lambdamid\otimes\Lambdamid\otimes\Lambdamid\otimes \Lambdamid$ in dashed purple, $\lambda(\Lambdamid\otimes\Lambdamid\otimes\Lambdamid\otimes \Lambdamid)$ in purple.}
    \label{autovalores 4 copias weak}
\end{figure}

Finally, taking advantage of our model, we have also investigated the behavior of the dynamical matrix's most negative eigenvalue for a range where $0.00 \leq \epsilon \leq 0.25$ in Fig.~\ref{autovalores 4 copias weak}. In this range, the dynamics is in the weak non-Markovian regime~\cite{nadja-NM,sabrina}. Differently from what we see in Fig.~\ref{autovalores 4 copias}, in this case, our protocol does not contribute to the distillation of non-Markovianity.  It seems that our protocol may make matters worse in this regime. Whenever we take many copies of the dynamical maps and send them through the displacement map $\lambda$, the least eigenvalue is non-negative, which signals that we end up with a Markovian dynamics. Note that this does not come from the fact that we are stacking up dynamical maps via the tensor product. For the case where we do take many copies, but there is no displacement map, the most negative eigenvalue is always negative for all $0.00 \leq \epsilon \leq 0.25$.

\section{Conclusion}
\label{sec:conclusion}

In this work, we proposed an operational method to distill the non-Markovianity of a quantum channel. In our framework, multiple copies of a channel are combined and mapped into a single channel (possibly) with a degree of non-Markovianity higher than the original channel, making use of a displacement map $\lambda$. Simply put, distill non-Markovianity from many copies of a given non-Markovian channel. To quantitatively analyze the distillation, we considered the increase in the distinguishability of two pure orthogonal states. In this contribution, using the collisional model, we worked out a concrete scenario where it was possible to achieve a distillation of non-Markovianity of quantum channels. Additionally, we showed that the distillation of 3 or more copies can provide further increases in non-Markovianity. 

Note that the scheme proposed here can be implemented with current technology. The idea of having many copies of the same quantum channel is very common in quantum communication schemes, especially in quantum repeater protocols. Mode and spatial multiplexing are essential for increasing transmission capacity in fibers. One natural experimental platform for multiple copies of the same channels is a multicore fiber. In this system, several cores are contained in the same cladding material in
a single fiber. Note that a single multicore fiber is surrounded by the same environment. Multicore fibers without noticeable crosstalk effects are already a reality (see for example \cite{Hayashi11} and more recently \cite{Carine20}). The coarse-graining map can be done in this scheme by multi-port beam splitters and measurements.

We also analyzed the behavior of the most negative eigenvalue of the dynamical matrix of the intermediate maps between the time $s$ and $t$ for the original dynamic (collisional model), as well as for the tensor products of the original dynamics (multiples copies of the channel) and the channel after the distillation. In so doing, we found that for our example and strong non-Markovianity, the most negative eigenvalue becomes more negative after the distillation—although it does not become as negative as the most negative eigenvalue of the tensor product. This does not prove that the more negative the most negative eigenvalue is, the more non-Markovian the dynamic is—as the measure of non-Markovianity is related to the area of the eigenvalue which is below zero, and in the cases presented here all areas diverge. We also observed that for the weak non-Markovian regime, the protocol destroyed non-Markovianity. The relation of different non-Markovian regimes (basically P-divisibility and non-P/CP-divisibility) and the success of the distillation protocol deserves further investigation.

As we have explored a particular scenario, one could investigate how to optimize the parameters in the problem to increase the gains of distinguishability in the distillation process—for instance, by finding different displacement maps that could increase it. Also, note that our work is also heavily based on the collisional model, so it is still open how our proposed distillation protocol behaves for other quantum dynamics, in particular, the study of quantum dynamics that are continuous in time, in contrast to the collisional model that is discreet in time, which could make for an interesting analysis.

\appendix
\section{Modified CP-Divisibility in the Proposed Protocol}
\label{sec:ModMarkovianityInProposedProtocol}

In this appendix, we elaborate on the notion of Modified CP-Divisibility and discuss that the operational protocol we have devised in our work does not contribute negatively to the distillation of non-Markovianity. We will show that our framework may well extract non-Markovianity from a Markovian dynamics, nonetheless that the extracted class of non-Markovianity is not fundamentally different from any Markovian dynamics---exactly as we have argued in the discussion around Eq.~\eqref{Eq.MotivatingModCPTPDiv}.  

In the argument leading to Def.~\ref{def:mod-divisibility}, we have seen that the composition $\lambda \circ \Theta_{t}$ may break the CP-divisibility of a CP-divisible family $\{ \Theta_{t} \}_{t}$. This is happens because for every $t \geq s \geq 0,$
\begin{align}
    \Theta_{t}^{\prime} = \lambda_{t} \circ \Theta_{t} = \lambda_{t} \circ V_{t,s} \circ \Theta_{s} = V_{t,s}^{\prime} \circ \Theta_{s},
\label{Eq.WhereCPTPDivFails_Appendix}
\end{align}
and there is no guarantee that we can remove $\lambda_{t}$ out of the equation. In this sense,  inputting a Markovian dynamics through $\lambda_{t}$ may result in a non-Markovian one---thus extracting non-Markovianity out of a Markovianity. Nonetheless, the only issue in the non-divisibility of $\{ \lambda_{t} \circ \Theta_{t}\}_{t}$ is the presence of the displacement map on the left-hand side in Eq.~\eqref{Eq.WhereCPTPDivFails_Appendix}. The functional form of that equation is not dissimilar to the one defining CP-divisible maps. It was exactly that similarity that motivated the definition of Modified CP-Divisibility---when $\lambda_{t}$ does not vary over time.

The major point missing from this discussion is the fact that in the operational quantum process that we have devised here, it is not the case that we are inputting $\Lambda_{t}$ through $\lambda$. What we do, instead, is send many copies of $\Lambda_{t}$ into $\lambda$. Put another way, with the notation of this appendix, $\Theta_{t} = \Lambda_{t}^{\otimes n}$, for a given integer $n > 0$ (see Fig.~\ref{boxes}). In principle, inputting many copies of $\Lambda_{t}$ through $\lambda$ could give rise to other forms of non-Markovianity---others different from the one expressed in Eq.~\ref{Eq.WhereCPTPDivFails_Appendix}. We emphasize that this is not the case. 

To see why this is the case, recall that if $\{\Lambda_{t}\}_{t}$ is CP-divisible, so $\{\Lambda_{t}^{ \otimes n}\}_{t}$ also is. To prove so, suppose that $\Lambda_{t} = V_{t,s} \circ \Lambda_{s}$, for all $t \geq s \geq 0$, then:
\begin{align}
    &\Lambda_{t}^{\otimes n}(\zeta) = (V_{t,s} \circ \Lambda_{s}) \otimes ... \otimes (V_{t,s} \circ \Lambda_{s})(\zeta) \nonumber \\
    & =  \sum_{i_1...i_N} \zeta_{i_1 ... i_N} (V_{t,s} \circ \Lambda_{s}) \otimes ... \otimes (V_{t,s} \circ \Lambda_{s})(\zeta_1 \otimes ... \otimes \zeta_N) \nonumber \\
    &=\sum_{i_1...i_N} \zeta_{i_1 ... i_N} V_{t,s}  (\Lambda_{s}(\zeta_1)) \otimes ... \otimes V_{t,s}  (\Lambda_{s}(\zeta_N)) \\
    &= \sum_{i_1...i_N} \zeta_{i_1 ... i_N} (V_{t,s} \otimes ... \otimes V_{t,s}) (\Lambda_{s}(\zeta_1) \otimes ... \otimes  \Lambda_{s}(\zeta_N)) \nonumber \\
    & = (V_{t,s} \otimes ... \otimes V_{t,s}) \left( \sum_{i_1...i_N} \zeta_{i_1 ... i_N} \Lambda_{s}(\zeta_1) \otimes ... \otimes  \Lambda_{s}(\zeta_N) \right) \nonumber \\
    &=  (V_{t,s} \otimes ... \otimes V_{t,s}) \left( \Lambda_{s}^{\otimes n} \left(  \sum_{i_1...i_N} \zeta_{i_1 ... i_N}  \zeta_1 \otimes ... \otimes  \zeta_N \right) \right)  \nonumber  \\
    &= V_{t,s}^{\otimes n } \circ \Lambda_{s}^{\otimes n} (\zeta). \nonumber
\end{align}
To conclude, it remains to show that the same functional form of Eq.~\eqref{Eq.WhereCPTPDivFails_Appendix} also holds true when we are considering many copies of a CP-Divisible family $\{\Lambda_{t}\}_{t}$. Again, suppose that $\Lambda_{t} = V_{t,s} \circ \Lambda_{s}$, for all $t \geq s \geq 0$, then:
\begin{align}
    \lambda \circ \Lambda_{t}^{\otimes n} &= \lambda \circ V_{t,s}^{\otimes n} \circ \Lambda_{s}^{\otimes n} 
     = (\lambda \circ V_{t,s}^{\otimes n}) \circ \Lambda_{s}^{\otimes n} 
\nonumber  \\
    & = V_{t,s}^{\prime} \circ \Lambda_{s}^{\otimes n}. 
    \label{Eq.ModCPCopies_Appendix}
\end{align}
In plain English, Eq.~\ref{Eq.ModCPCopies_Appendix} says that our protocol can extract non-Markovianity out of a Markovian dynamics, nonetheless, this non-Markovianity has a special form---it is that one we dubbed \emph{Modified Markovian}. All in all, the number of copies we decide to work with at the beginning of the protocol does not bring any qualitative impact on the creation of non-Markovianity out of a Markovian dynamics. Any non-Markovianity created (out of a Markovian dynamics) is always of the same type.

\section{The coarse-graining displacement map without a non-markovian dynamic}
\label{sec:coarse-graining-map}

As we mentioned in the main text, in this Appendix, we analyze the action of the coarse-graining displacement map for the case of two copies of the system $\rho$, without the presence of the non-Markovian time evolution. To do so, we apply $\lambda$ on two copies of the state $\rho$, represented in the Bloch sphere by the vector $\Vec{r}=r( \sin(\theta)\cos(\phi), \sin(\theta)\sin(\phi),\cos(\theta))$. The resulting state $\rho'$ after applying the coarse-graining map $\lambda$ in two copies of the state $\rho$ is given by

\begin{strip}
\begin{align}
    \label{eq:lambda bloch sphere}
\rho' = \lambda (\rho\otimes\rho) 
 = \left(
\begin{array}{cc}
 \frac{1}{4} (1+r \cos(\theta))^2 & 0 \\
 0 & \frac{1}{4} (1-r \cos(\theta))^2+\frac{1}{2} (1-r^2 \cos^2(\theta))
\end{array}
\right)
\end{align}
\end{strip}


Notice that the the off-diagonal terms are killed off, for any state $\rho$, and the non-zero component depends exclusively on $r$ and $\theta$. 
We want to study the case of pure orthogonal states $\rho_1$ and $\rho_2$ given by Equations~\eqref{eq:pure_orthogonal_1} and ~\eqref{eq:pure_orthogonal_2}, whose Bloch vectors are $\Vec{r_1}=( \sin(\theta)\cos(\phi), \sin(\theta)\sin(\phi),\cos(\theta))$ and $\Vec{r_2} = ( -\sin(\theta)\cos(\phi), -\sin(\theta)\sin(\phi), - \cos(\theta))$. We can then evaluate the final distinguishability, which gives us:

\begin{equation}\label{eq:D bloch sphere}
    D(\rho'_1, \rho'_2)=\frac{1}{2}\left\Vert \rho'_2-\rho'_1 \right\Vert_1=|\cos(\theta)|  
\end{equation}

So, for pure orthogonal states, the coarse-graining map alone does not increase the distinguishability. Consequently, in the example we address in our framework, the increase in distinguishability comes from the use of the coarse-graning map associated with the non-Markovian dynamic.

Nonetheless, in the general case the situation may be different. For example, if we take two states with $r_1=1$, $r_2=0.5$, $\theta_1 = \theta_2 = 0$, and $\phi_1= \phi_2 = 0$. After applying the coarse-graining map, we obtain $r'_1=1$ and $r'_2=1/8$, which will lead to an increase in the distinguishability, i.e. $ D(\rho_1, \rho_2) = 0.25 < D(\rho'_1, \rho'_2) = 0.44$. So, for general states, the distinguishability can increase with the application of the coarse-graining displacement map.

\section*{Acknowledgements}

The authors thank L. Aolita for the initial discussions and his helpful suggestions on this work. N.K.B. acknowledges financial support from CNPq Brazil (Universal Grant No. 406499/2021-7) and FAPESP (Grant 2021/06035-0). N.K.B. is part of the Brazilian National Institute for Quantum Information (INCT Grant 465469/2014-0). C.D. is supported by an Engineering and Physical Sciences Research Council (EPSRC) grant, Grant No. EP/V002732/1. C.D. is also supported by a visiting professorship from CAPES. He is grateful for the hospitality of the Departamento de Física da Universidade Federal de Pernambuco. During part of the execution of this project, T.M.D.A. was also supported by a scholarship from CNPq. This study was financed in part by the Coordenação de Aperfeiçoamento de Pessoal de Nível Superior – Brasil (CAPES) – Finance Code 001. This work is licensed under a Creative Commons Attribution-NonCommercial-NoDerivs 4.0 International License.


\begin{thebibliography}{99}

\bibitem{breuer} H. P. Breuer and F. Petruccione, \textsl{The Theory of Open Quantum Systems}. Oxford University Press, New York, 2007.

\bibitem{Rivas14} Á. Rivas, S. F. Huelga, and M. B. Plenio, Rep. Prog. Phys. \textbf{77}, 094001 (2014).

\bibitem{Vega17} I. de Vega and D. Alonso, Rev. Mod. Phys. \textbf{89}, 015001 (2017).

\bibitem{matsuzaki} Y. Matsuzaki, S. C. Benjamin, and J. Fitzsimons, Phys. Rev. A \textbf{84}, 012103 (2011).

\bibitem{verstraete} F. Verstraete, M. M. Wolf, and J. I. Cirac, Nature Phys. \textbf{5}, 633 (2009).

\bibitem{vasile} R. Vasile, S. Olivares, M. G. A. Paris, and S. Maniscalco, Phys. Rev. A \textbf{83}, 042321 (2011).

\bibitem{chin} A. W. Chin, S. F. Huelga, and M. B. Plenio, Phys. Rev. Lett. \textbf{109}, 233601 (2012).

\bibitem{bogna} B. Bylicka, D. Chruscinski, and S. Maniscalco, Sci Rep \textbf{4}, 5720 (2014).

\bibitem{wolf1}  M. M. Wolf and J. I. Cirac, Commun. Math. Phys. \textbf{279}, 147 (2008).

\bibitem{Alicki06} R. Alicki and K. Lendi, \textit{ Quantum Dynamical Semigroups
and Applications}. Springer, Berlin, (2006).

\bibitem{Milz19} S. Milz, M. S. Kim, F. A. Pollock, and K. Modi, Phys. Rev. Lett. \textbf{123}, 040401 (2019).

\bibitem{Thomas18} G. Thomas, N. Siddharth, S. Banerjee, and S. Ghosh, Phys. Rev. E \textbf{97}, 062108 (2018).

\bibitem{Reich15} D. M. Reich, N. Katz, and C. P. Koch, Sci. Rep. 5, 12430 (2015).

\bibitem{elsi} E.-M. Laine, H.-P. Breuer, and J. Piilo, Sci. Rep. \textbf{4}, 4620 (2014).

\bibitem{berk2021} G. D. Berk, A. J. P. Garner, B. Yadin, K. Modi, and F. A. Pollock. Quantum \textbf{5}, 435 (2021).

\bibitem{bennett} C. H. Bennett, G. Brassard, S. Popescu, B. Schumacher, J. A. Smolin, and W. K. Wootters, Phys. Rev. Lett. \textbf{76}, 722 (1996).

\bibitem{dur} W. D\"ur, H.-J. Briegel, J. I. Cirac,  P. Zoller, Phys. Rev. A, \textbf{59}, 169 (1999).

\bibitem{gisin} N. Gisin and R. Thew. Nature Photonics, \textbf{1}, 165 (2007).

\bibitem{Wilde19a} X. Wang and M. M. Wilde, Phys. Rev. Research \textbf{1}, 033169 (2019).

\bibitem{Wilde19b} X. Wang and M. M. Wilde, Phys. Rev. Research \textbf{1}, 033170 (2019).

\bibitem{Regula} B. Regula and R. Takagi, Nature Communications \textbf{12}, 4411 (2021).

\bibitem{Maity} A. G. Maity, S. Bhattacharya, arXiv:2206.04524 (2022).

\bibitem{Bhattacharya20} S. Bhattacharya, B. Bhattacharya, and A. S. Majumdar, J. Phys. A: Math. Theor. \textbf{53} 335301 (2020).

\bibitem{Anand} N. Anand, T. A. Brun, arXiv:1903.03880 (2019).

\bibitem{Berk2} G. D. Berk, S. Milz, F. A. Pollock, K. Modi, arXiv:2110.02613 (2021).

\bibitem{Takagi} Ryuji Takagi, Xiao Yuan, Bartosz Regula, and Mile Gu, Phys. Rev. A \textbf{109}, 022403 (2024).

\bibitem{Choi} M. Choi, Linear Algebra and Its Applications, \textbf{10}, 285 (1975).

\bibitem{Kraus} K. Kraus, \textsl{States Effects and Operations: Fundamental Notions of Quantum Theory} (Springer Verlag, 1983).

\bibitem{rivas}  A. Rivas,  S. F. Huelga and M. B. Plenio, Phys. Rev. Lett. \textbf{105}, 050403 (2010).

\bibitem{blp} H.-P. Breuer, E.-M. Laine, and J. Piilo, Phys. Rev. Lett. \textbf{103}, 210401 (2009).

\bibitem{salvail} L. Salvail. \textit{The Search for the Holy Grail in Quantum Cryptography}. In: Damgård, I.B. (eds) Lectures on Data Security. EEF School 1998. Lecture Notes in Computer Science, vol 1561. Springer, Berlin, Heidelberg (1999).

\bibitem{Vedral10} V. Vedral, Found. Phys. \textbf{40}, 1141 (2010).

\bibitem{Chris-coarse} C. Duarte, G. D. Carvalho, N. K. Bernardes, and F. de Melo, Phys. Rev. A \textbf{96}, 032113 (2017).

\bibitem{Eberly07} T. Yu, J.H. Eberly, Quantum Information and Computation \textbf{7}, 459 (2007).

\bibitem{nadja-NM} N. K. Bernardes, et al. Sci. Rep. \textbf{5}, 17520 (2015).

\bibitem{sabrina} D. Chruscinski and S. Maniscalco, Phys. Rev. Lett. \textbf{112}, 120404 (2014).

\bibitem{Dariusz15} D. Chruscinski, F. A. Wudarski, Phys. Rev. A \textbf{91}, 012104 (2015).

\bibitem{CICCARELLO20221} F. Ciccarello, S. Lorenzo, V. Giovannetti and G. Massimo Palma, Physics Reports, \textbf{954}, 1 (2022).

\bibitem{Rybar} Tomáš Rybár et al, J. Phys. B: At. Mol. Opt. Phys. \textbf{45}, 154006 (2012).

\bibitem{Wissmann} S. Wi\ss{}mann, A. Karlsson, E.-M. Laine, J. Piilo, and H.-P. Breuer, Phys. Rev. A \textbf{86}, 062108 (2012).

\bibitem{Jamiokowski72} A.Jamiołkowski, Reports on Mathematical Physics \textbf{3} , 275 (1972).

\bibitem{Bengtsson06} I. Bengtsson and K. Zyczkowski, Geometry of Quantum States: An Introduction to Quantum Entanglement (Cambridge
University Press, Cambridge, UK, 2006), Sec. 10.3.

\bibitem{Hayashi11} T. Hayashi, T. Taru, O. Shimakawa, T. Sasaki, and E. Sasaoka, Opt. Express \textbf{19}, 16576 (2011).

 \bibitem{Carine20} J. Carine, G. Cañas, P. Skrzypczyk, I. Šupić, N. Guerrero, T. Garcia, L. Pereira, M. A. S. Prosser, G. B. Xavier, A. Delgado, S. P. Walborn, D. Cavalcanti, and G. Lima, Optica \textbf{7}, 542 (2020).

\end{thebibliography}


\end{document}